\documentclass[a4paper,10pt]{article} 

\newcommand\draft[1]{}
\newcommand\release[1]{#1}
\newcommand\bigone[1]{}
\newcommand\smallone[1]{#1}

\newcommand{\cD}{{\cal D}}

\draft{\usepackage{easy-todo}}
\draft{\usepackage[notref,notcite]{showkeys}}

\usepackage{graphicx}

\usepackage{amsfonts}
\usepackage{amssymb}
\usepackage{amsmath}
\usepackage{amsthm}
\usepackage{latexsym}

\bigone{\newtheorem{thm}{Theorem}[chapter]}
\smallone{\newtheorem{thm}{Theorem}}
\newtheorem{lem}[thm]{Lemma}
\newtheorem{prp}[thm]{Proposition}

\newtheorem{clm}[thm]{Claim}

\theoremstyle{definition}

\newtheorem{defn}[thm]{Definition}

\newcommand{\refthm}[1]{Theorem~\ref{thm:#1}}
\newcommand{\reflem}[1]{Lemma~\ref{lem:#1}}
\newcommand{\refprp}[1]{Proposition~\ref{prp:#1}}

\newcommand{\refclm}[1]{Claim~\ref{clm:#1}}

\newcommand{\refdefn}[1]{Definition~\ref{defn:#1}}

\newcommand{\refsec}[1]{Section~\ref{sec:#1}}
\newcommand{\reffig}[1]{Figure~\ref{fig:#1}}
\newcommand{\refeqn}[1]{(\ref{eqn:#1})}

\newcommand{\pfstart}{\begin{proof}} 
\newcommand{\pfsketch}{\begin{proof}[Proof sketch]}
\newcommand{\pfend}{\end{proof}} 
\newcommand{\itemstart}{\begin{itemize}\itemsep0pt}
\newcommand{\itemend}{\end{itemize}}
\newcommand{\descrstart}{\begin{description}\itemsep0pt}
\newcommand{\descrend}{\end{description}}
\newcommand{\enumstart}{\begin{enumerate}\itemsep0pt}
\newcommand{\enumend}{\end{enumerate}}

\newcommand{\C}{{\mathbb{C}}}

\newcommand{\R}{{\mathbb{R}}}
\newcommand{\Z}{{\mathbb{Z}}}

\newcommand{\ii}{\mathsf{i}}
\newcommand{\ee}{\mathsf{e}}

\usepackage{xspace} 
\newcommand{\etal}{{\em et al.}\xspace}

\newcommand{\ignore}[1]{}

\def \s[#1]{\left(#1\right)}
\def \sA[#1]{\bigl(#1\bigr)}
\def \sB[#1]{\Bigl(#1\Bigr)}
\def \sC[#1]{\biggl(#1\biggr)}
\def \sD[#1]{\Biggl(#1\Biggr)}

\def \sk[#1]{\left[#1\right]}
\def \skA[#1]{\bigl[#1\bigr]}
\def \skB[#1]{\Bigl[#1\Bigr]}
\def \skC[#1]{\biggl[#1\biggr]}
\def \skD[#1]{\Biggl[#1\Biggr]}

\def \abs|#1|{\left| #1\right|}
\def \absA|#1|{\bigl|#1\bigr|}
\def \absB|#1|{\Bigl|#1\Bigr|}
\def \absC|#1|{\biggl|#1\biggr|}
\def \absD|#1|{\Biggl|#1\Biggr|}

\def \norm|#1|{\left\| #1\right\|}
\def \normA|#1|{\bigl\| #1\bigr\|}
\def \normB|#1|{\Bigl\| #1\Bigr\|}
\def \normC|#1|{\biggl\| #1\biggr\|}
\def \normD|#1|{\Biggl\| #1\Biggr\|}
\def \normS|#1|{\| #1\|}

\def \normFrob|#1|{\norm|#1|_{\mathrm{F}}}

\def \sfig#1{\left\{#1\right\}}
\def \elem[#1]{[\![#1]\!]}

\def \ket#1|#2>{|#2\rangle_{\mathcal{#1}}}

\def \ip<#1,#2>{\langle #1, #2\rangle}
\def \ipA<#1,#2>{\bigl\langle #1, #2\bigr\rangle}
\def \ipB<#1,#2>{\Bigl\langle #1, #2\Bigr\rangle}
\def \ipC<#1,#2>{\biggl\langle #1, #2\biggr\rangle}
\def \ipD<#1,#2>{\Biggl\langle #1, #2\Biggr\rangle}

\newcommand{\polylog}{\mathop{\mathrm{polylog}}}

\def\mycommand#1#2{%
\draft{\marginpar{\fbox{\sf{#1:} $#2$}}}%
\expandafter\newcommand \csname#1\endcsname {#2}%
}

\usepackage{fullpage}
\usepackage{url}
\release{\usepackage{hyperref}}

\newcommand{\Adv}{\mathrm{Adv}^\pm}

\input{xy}
\xyoption{all}
\release{
\xyoption{dvips}
}

\title{On the Power of Non-Adaptive Learning Graphs}
\author{
Aleksandrs Belovs\thanks{Faculty of Computing, University of Latvia, \protect\url{stiboh@gmail.com}}
\and
Ansis Rosmanis\thanks{David R. Cheriton School of Computer Science and
Institute for Quantum Computing, University of Waterloo, \protect\url{arosmanis@uwaterloo.ca}}
}
\date{}

\begin{document}
\maketitle

\begin{abstract}
We introduce a notion of the quantum query complexity of a certificate structure. This is a formalisation of a well-known observation that many quantum query algorithms only require the knowledge of the disposition of possible certificates in the input string, not the precise values therein.  

Next, we derive a dual formulation of the complexity of a non-adaptive learning graph, and use it to show that non-adaptive learning graphs are tight for all certificate structures.  By this, we mean that there exists a function possessing the certificate structure and such that a learning graph gives an optimal quantum query algorithm for it.  

For a special case of certificate structures generated by certificates of bounded size, we construct a relatively general class of functions having this property.  The construction is based on orthogonal arrays, and generalizes the quantum query lower bound for the $k$-sum problem derived recently~\cite{spalek:kSumLower}.

Finally, we use these results to show that the learning graph for the triangle problem from Ref.~\cite{lee:learningTriangle} is almost optimal in these settings.  This also gives a quantum query lower bound for the triangle-sum problem.
\end{abstract}

\section{Introduction}

\mycommand{cert}{{\cal C}}
\mycommand{marked}{M}

Determining the amount of computational resources required to solve a computational problem is one of the main problems in theoretical computer science.  At the current stage of knowledge, however, this task seems far out of reach for many problems.  In this case, it is possible to analyse the complexity of the problem under some simplifying assumptions.

One of such assumptions is exhibited by the query model.  In this model, it is assumed that all computational resources except accessing the input string are free of charge.  (For a detailed description of the model, including our case of interest---quantum query complexity, refer to~\cite{buhrman:querySurvey}.)  Under this assumption, it is possible to prove some tight bounds.  In particular, a relatively simple semidefinite program (SDP) was constructed, yielding a tight estimate for the quantum query complexity of any function.  This is the adversary bound, we describe in \refsec{adversary}.

Unfortunately, for many functions, even this SDP is too hard to solve.  In this paper, we investigate a possibility of constructing an even simpler optimization problem under further simplifying assumptions.  Our assumptions are motivated by the class of algorithms based on quantum walks.  A popular framework for the development of such algorithms~\cite{magniez:walkSearch} includes a black-box {\em checking} subroutine that, given the information gathered during the walk, signals if this information is enough to accept the input string.  In many cases, the precise content of the gathered information is not relevant for the implementation of the quantum walk, what matters are the possible locations of these pieces of information.  We formalise this by the following definition. 

In the definition, we use the following notations.  If $m$ and $n$ are positive integers, we use $[n]$ to denote the set $\{1,2,\dots,n\}$, and $[m,n]$ to denote the set $\{m,m+1,\dots,n\}$.  
Also, for a sequence $x=(x_i)\in [q]^n$ and $S\subseteq [n]$, let $x_S\in [q]^S$ denote the projection of $x$ on $S$, i.e., the sequence $(x_{s_1},\dots,x_{s_\ell})$ indexed by the elements $s_1,\dots,s_\ell$ of $S$.

\begin{defn}[Certificate Structure]
A {\em certificate structure} $\cert$ on $n$ variables is a collection of non-empty subsets of $2^{[n]}$ with each subset closed under taking supersets.  We say a function $f\colon \cD\to\{0,1\}$ with $\cD\subseteq [q]^n$ {\em has} certificate structure $\cert$ if, for every $x\in f^{-1}(1)$, one can find $\marked\in\cert$ such that
\[
\forall S\in\marked\; \forall z\in\cD\colon z_S = x_S \Longrightarrow f(z)=1.
\]
\end{defn}

We are interested in quantum algorithms performing equally well for any function with a fixed certificate structure.  Some examples of such algorithms are given in \refsec{examples}.  More formally, define the {\em quantum query complexity of a certificate structure} as the maximum quantum query complexity over all functions possessing this certificate structure.

A recently developed computation model of a (non-adaptive) learning graph~\cite{belovs:learning} relies on the certificate structure of the function by definition.  This suggests to define the {\em learning graph complexity of a certificate structure} as the minimum complexity of a non-adaptive learning graph computing a function (hence, any function) with this certificate structure.  Since a learning graph can be transformed into a quantum query algorithm with the same complexity, the learning graph complexity of a certificate structure is an upper bound on its quantum query complexity.  In this paper, we prove that these two complexities are actually equal up to a constant factor.
\begin{thm}
\label{thm:certificates}
For any certificate structure, its quantum query and learning graph complexities differ by at most a constant multiplicative factor.
\end{thm}

This means that any quantum query algorithm willing to perform better than the best learning graph has to take the values of the variables into account on the earlier stages of the algorithm.
\mycommand{certM}{{A_\marked}}
Although \refthm{certificates} is a very general result, it is unsatisfactory in the sense that the function having the required quantum query complexity is rather artificial, and the size of the alphabet is astronomical.  However, for a special case of certificates structures we are about to define, it is possible to construct a relatively natural problem with a modestly-sized alphabet having high quantum query complexity.

\begin{defn}[Boundedly Generated Certificate Structure]
\label{defn:boundedly}
We say that a certificate structure $\cert$ is {\em boundedly generated} if, for any $\marked\in\cert$, one can find a subset $\certM\subseteq [n]$ such that $|\certM| = O(1)$, and $S\in M$ if and only if $S\supseteq \certM$. 
\end{defn}

%In order to define the function with a high complexity, we first have to introduce the following special case of a well-studied combinatorial object. 

\begin{defn}[Orthogonal Array]\label{defn:orthogonalArray}
Assume $T$ is a subset of $[q]^k$.  We say that $T$ is an \emph{orthogonal array} over alphabet $[q]$ iff, for every index $i \in [k]$ and for every sequence $x_1,\dots, x_{i-1}, x_{i+1},\dots, x_k$ of elements in $[q]$, there exist exactly $|T|/q^{k-1}$ choices of $x_i \in [q]$ such that $(x_1, \dots, x_k) \in T$.  We call $|T|$ the {\em size} of the array, and $k$---its {\em length}.
(Compared to a standard definition of orthogonal arrays (cf.~\cite{hedayat:orthogonal}), we always require that the so-called strength of the array equals $k-1$.)
\end{defn}

\begin{thm}
\label{thm:boundedly}
Assume a certificate structure $\cert$ is boundedly generated, and let $\certM$ be like in \refdefn{boundedly}.  Assume the alphabet is $[q]$ for some $q\ge 2|\cert|$, and each $\certM$ is equipped with an orthogonal array $T_\marked$ over alphabet $[q]$ of length $|\certM|$ and size $q^{|\certM|-1}$.  Consider a function $f\colon [q]^n\to\{0,1\}$ defined by $f(x)=1$ iff there exists $\marked\in\cert$ such that $x_{\certM} \in T_\marked$.  Then, the quantum query complexity of $f$ is at least a constant times the learning graph complexity of $\cert$.
\end{thm}

For example, for a boundedly generated certificate structure $\cert$, one can define the corresponding {\em sum} problem: Given $x\in[q]^n$, detect whether there exists $M\in\cert$ such that $\sum_{j\in\certM} x_j\equiv 0\pmod{q}$.  If $q\ge 2|\cert|$, \refthm{boundedly} implies that the quantum query complexity of this problem is at least a constant times the learning graph complexity of $\cert$.

\refthm{boundedly} is a generalization of the lower bound for the $k$-sum problem from Ref.~\cite{spalek:kSumLower}, and provides additional intuition on the construction, by linking it to learning graphs.  Much of the discussion in Ref.~\cite{spalek:kSumLower} applies here as well.

Let us briefly comment on organization of the paper.  In \refsec{examples}, we give some examples of certificate structures, inspired by known computational problems.  In \refsec{learning}, we derive a dual formulation of the complexity of a non-adaptive learning graph.  In \refsec{applications}, we apply this dual formulation to give lower bounds on the learning graph complexity of the certificate structures from \refsec{examples}.  We demonstrate that transition to the learning graph complexity indeed simplifies the problem by obtaining an almost optimal $\widetilde{\Omega}(n^{9/7})$ lower bound for the triangle certificate structure, whereas nothing better than trivial $\Omega(n)$ is known for the original triangle problem.  Finally, in \refsec{lower}, we prove both Theorem~\ref{thm:certificates} and~\ref{thm:boundedly}.

\section{Examples of Certificate Structures}
\label{sec:examples}
We defined the certificate structure notion in the introduction.  Actually, many existing quantum algorithms, implicitly or explicitly, work in these settings.  In this section, we recall some of these algorithms and define the corresponding certificate structures.  In \refsec{applications}, we consider their learning graph complexities.

The most celebrated examples of such algorithms are demonstrated by Grover's search algorithm~\cite{grover:search}, and Ambainis' algorithm for element distinctness and $k$-distinctness~\cite{ambainis:distinctness}.  As first noticed by Childs and Eisenberg~\cite{childs:subsetFinding}, Ambainis' algorithm can be applied for finding any subset of size $k$.  In other words, it works for any function having the following certificate structure:

\begin{defn}
The {\em $k$-subset certificate structure} $\cert$ on $n$ elements with $k=O(1)$ is defined as follows.  It has ${n\choose k}$ elements, and, for each subset $A\subseteq[n]$ of size $k$, there exists unique $M\in\cert$ such that $S\in M$ if and only if $A\subseteq S\subseteq [n]$.
\end{defn}

In the same paper, Childs and Eisenberg conjectured that Ambainis' algorithm is optimal for the $k$-sum problem.  \refthm{boundedly} can be seen as a strong generalization of this conjecture (as Ambainis' algorithm can be implemented as a learning graph).  

Another well-known quantum-walk-based algorithm~\cite{magniez:triangle} (implicitly) solves any function with the following certificate structure:

\begin{defn}
The {\em triangle certificate structure} $\cert$ on $n$ vertices is a certificate structure on ${n\choose 2}$ variables defined as follows.  Assume that the variables are labelled as $x_{ij}$ where $1\le i<j\le n$.  The certificate structure has ${n\choose 3}$ elements, and, for every triple $1\le a<b<c\le n$, there exists unique $M\in\cert$ such that $S\in M$ if and only if $S\supseteq \{ab,bc,ac\}$.  (Note that, for this certificate structure, the letter $n$, that customary denotes the number of input variables, is used to denote the number of vertices.  This is a standard notation, and we hope it will not cause much confusion.)
\end{defn}

Originally, the algorithm in Ref.~\cite{magniez:triangle} dealt with the {\em triangle problem}: All $x_{ij}$ are Boolean, and the condition on $f(x)=1$ is that $x_{ab}=x_{ac}=x_{bc}=1$ for some $M$.  The quantum walk algorithm for this certificate structure was lately superseded by an algorithm based on learning graphs~\cite{lee:learningTriangle}.  We will show in \refsec{applications} that this learning graph is esentially optimal.  

Both $k$-subset and triangle certificate structures are boundedly generated.  We also consider some examples of certificate structures that are not.  Recall the {\em collision problem}~\cite{brassard:collision}.  Given an input string $x\in[q]^{2n}$, the task is to distinguish two cases.  In the negative case, all input variables are distinct.  In the positive case, there exists a decomposition of the input variables $[2n]=\{a_1,b_1\}\sqcup\{a_2,b_2\}\sqcup\cdots\sqcup\{a_n,b_n\}$ into $n$ pairs such that $x_{a_i}=x_{b_i}$ for all $i\in[n]$, but $x_{a_i}\ne x_{a_j}$ for all $i\ne j$.  The {\em set equality problem} is defined similarly, with an additional promise that, in the positive case, $a_i\in [n]$ and $b_i\in[n+1,2n]$ for all $i$.  Finally, the {\em hidden shift problem} is defined like the set equality problem with an additional promise that, in the positive case, there exists $d\in[n]$ such that $b_i = n+1+((a_i+d)\bmod n)$ for all $i\in[n]$.  Inspired by these problems, we define the following certificate structures.

\begin{defn}
\label{defn:collision}
Each of the following certificate structures is defined on $2n$ input variables.  In the {\em collision certificate structure}, there is unique $M$ for each decomposition $[2n]=\{a_1,b_1\}\sqcup\{a_2,b_2\}\sqcup\cdots\sqcup\{a_n,b_n\}$, and $S\in M$ if and only if $S\supseteq\{a_i,b_i\}$ for some $i\in[n]$.  The {\em set equality certificate structure} contains only those $M$ from the collision certificate structure that correspond to decompositions with $a_i\in[n]$ and $b_i\in[n+1,2n]$ for all $i$.  Finally, the {\em hidden shift certificate structure} contains only those $M$ from the set equality certificate structure that correspond to decompositions such that $d\in[n]$ exists with the property $b_i = n+1+((a_i+d)\bmod n)$ for all $i\in[n]$.
\end{defn}

All certificates structure from~\refdefn{collision} are not boundedly generated.  The algorithm for the collision problem from Ref.~\cite{brassard:collision} actually solves any function possessing the collision certificate structure in $O(n^{1/3})$ quantum queries, and it is tight~\cite{shi:collisionLower}.  Clearly, the same algorithm is applicable for the set equality and hidden shift certificate structures.  The situation with the hidden shift problem is more interesting.  This problem reduces to the hidden subgroup problem in the dihedral group~\cite{kuperberg:dihedral}, and the latter has logarithmic query complexity~\cite{ettinger:hspQuery}.  Unlike other algorithms in this section, the latter one is not, in general, applicable to any function with the hidden shift certificate structure.

\section{Learning Graph Complexity}
\label{sec:learning}
\mycommand{arcs}{{\cal E}}
\mycommand{source}{\mathrm{s}}
\mycommand{target}{\mathrm{t}}

In this section, we recall the definition of a non-adaptive learning graph from Ref.~\cite{belovs:learning}, and derive its dual formulation.  Although more general concepts of learning graphs were introduced~\cite{lee:learningKdistPrior, belovs:learningKDist, gavinsky:graphCollision}, the non-adaptive version was used extensively~\cite{zhu:learning, lee:learningSubgraphs, lee:learningTriangle}, mostly because of its simplicity.  Hence, it is important to understand its limitations.

Let $\arcs$ by the set of pairs $(S,S')$ of subsets of $[n]$ such that $S'=S\cup\{j\}$ for some $j\notin S$.  This set is known as the {\em set of arcs} of a learning graph on $n$ variables.  For $e=(S,S')\in\arcs$, let $\source(e)=S$ and $\target(e)=S'$.

\begin{defn}
\label{defn:main}
The learning graph complexity of a certificate structure $\cert$ on $n$ variables is equal to the optimal value of the following two optimization problems
\begin{subequations}
\label{eqn:learningPrimal}
\begin{alignat}{3}
 &{\mbox{\rm minimize }} &\quad& \sqrt{\sum\nolimits_{e\in\arcs} w_e} \label{eqn:flowObjective} \\ 
 &{\mbox{\rm subject to }} && \sum\nolimits_{e\in\arcs} \frac{p_e(\marked)^2}{w_e}\le 1 &\quad&\text{\rm for all $\marked\in\cert$;} \label{eqn:flowValue}\\
 &&& \sum_{e\in\arcs\colon \target(e)= S} p_e(\marked) = \sum_{e\in\arcs\colon \source(e)= S} p_e(\marked) &&\text{\rm for all $\marked\in\cert$ and $S\in 2^{[n]}\setminus(M\cup\{\emptyset\})$;} \label{eqn:flowCond1} \\
 &&& \sum\nolimits_{e\in\arcs\colon \source(e) = \emptyset} p_e(\marked) = 1 &&\text{\rm for all $\marked\in\cert$;}\label{eqn:flowCond2}\\
 &&& p_e(M)\in\R,\quad w_e\ge 0 && \text{\rm for all $e\in\arcs$ and $M\in\cert$;}
\end{alignat}
\end{subequations}
(here, $0/0$ in~\refeqn{flowValue} is defined to be 0), and
\begin{subequations}
\label{eqn:learningDual}
\begin{alignat}{3}
 &{\mbox{\rm maximize }} &\quad& \sqrt{\sum\nolimits_{\marked\in\cert} \alpha_\emptyset(\marked)^2} \label{eqn:alphaObjective} \\ 
 &{\mbox{\rm subject to }} && \sum\nolimits_{\marked\in\cert} \s[\alpha_{\source(e)}(\marked) - \alpha_{\target(e)}(\marked)\strut]^2\le 1 &\quad&\text{\rm for all $e\in\arcs$;} \label{eqn:alphaOne} \\
 &&& \alpha_S(\marked) = 0 && \text{\rm whenever $S\in \marked$;} \label{eqn:alphaZero}  \\
 &&& \alpha_S(\marked)\in\R && \text{\rm for all $S\subseteq[n]$ and $M\in\cert$.}
\end{alignat}
\end{subequations}
\end{defn}

Eq.~\refeqn{learningPrimal} is a restatement of the definition of a non-adaptive learning graph from Ref.~\cite{belovs:learning}.  (In Ref.~\cite{belovs:learning}, the complexity was defined as the minimum of~\refeqn{flowObjective} and the maximum of the left hand side of~\refeqn{flowValue} over all $M$.  The current formulation can be obtained by rescaling all $p_e(M)$ by the same factor.  See also Footnote 1 in Ref.~\cite{belovs:learningKDist}.)  The second expression~\refeqn{learningDual} is a new one, and requires a proof.  

\pfstart[Proof of the equivalence of~\refeqn{learningPrimal} and~\refeqn{learningDual}]
The equivalence is obtained by duality.  We use basic convex duality~\cite[Chapter 5]{boyd:convex}.
First of all, we consider both programs with their objective values~\refeqn{flowObjective} and~\refeqn{alphaObjective} squared.  With this change, Eq.~\refeqn{learningPrimal} becomes a convex program  (for the convexity of~\refeqn{flowValue}, see Ref.~\cite[Section 3.1.5]{boyd:convex}).  The program is strictly feasible.  Indeed, it is easy to see that~\refeqn{flowCond1} and~\refeqn{flowCond2} are feasible.  To assure strong feasibility in~\refeqn{flowValue}, it is enough to take $w_e$ large enough.  Hence, by Slater's condition, the optimal values of~\refeqn{learningPrimal} and its dual are equal.  Let us calculate the dual.  The Lagrangian of~\refeqn{learningPrimal} is as follows
\begin{multline}
\label{eqn:lagrangian}
\sum_{e\in\arcs}w_e + \sum_{M\in\cert}\mu_M\s[\sum_{e\in\arcs} \frac{p_e(\marked)^2}{w_e}-1] \\
+\sum_{\substack{M\in\cert,\; S\subseteq[n]\\S\ne\emptyset,\; S\notin M}} \nu_{M,S}
\sD[\sum_{\substack{e\in\arcs\\\target(e)=S}} p_e(M) - \sum_{\substack{e\in\arcs\\\source(e)=S}} p_e(M)] +
\sum_{M\in\cert} \nu_{M,\emptyset}\sD[1-\sum_{\substack{e\in\arcs\\\source(e)=\emptyset}} p_e(M)].
\end{multline}
Here $\mu_M\ge 0$, and $\nu_{M,S}$ are arbitrary.  Let us first minimize over $p_e(M)$.  Each $p_e(M)$ appears three times in~\refeqn{lagrangian} with the following coefficients:
\[
p_e(M)^2\frac{\mu_M}{w_e} + p_e(M)\sA[\nu_{M,\target(e)}-\nu_{M,\source(e)}],
\]
where we assume $\nu_{M,S}=0$ for all $S\in M$.  The minimum of this expression clearly is 
\[
-\frac{w_e}{4\mu_M} \sA[\nu_{M,\target(e)}-\nu_{M,\source(e)}]^2.
\]
Plugging this into~\refeqn{lagrangian} yields
\begin{equation}
\label{eqn:lagrangian2}
\sum_{M\in\cert} (\nu_{M,\emptyset}-\mu_M) + 
\sum_{e\in\arcs} w_e\s[1 - \sum_{M\in\cert} \frac{\sA[\nu_{M,\target(e)}-\nu_{M,\source(e)}]^2}{4\mu_M}].
\end{equation}
Define $\alpha_S(M)$ as $\nu_{M,S}/(2\sqrt{\mu_M})$.  Minimizing~\refeqn{lagrangian2} over $w_e$, the second term disappears if condition~\refeqn{alphaOne} is satisfied.  The first term is
\[
\sum_{M\in\cert} (2\sqrt{\mu_M}\alpha_\emptyset(M) - \mu_M).
\]
We can also maximize over $\mu_M$, that gives the square of~\refeqn{alphaObjective}.
\pfend

We have the following result:
\begin{thm}[\cite{belovs:learning, lee:learningKdistPrior}]
The quantum query complexity of a certificate structure is at most a constant times its learning graph complexity.
\end{thm}

In \refsec{lower}, we prove the reverse statement for all certificate structures.

\section{Examples of Application}
\label{sec:applications}
\mycommand{alphaz}{\alpha_\emptyset}
In this section, we construct feasible solutions to the dual formulation of the learning graph complexity~\refeqn{learningDual} for the certificate structures from \refsec{examples}.  Their objective values match the objective values of feasible solutions to the corresponding primal formulations~\refeqn{learningPrimal} that were obtained previously.

\begin{prp}
\label{prp:ksum}
The learning graph complexity (and, hence, the quantum query complexity) of the $k$-subset certificate structure is $\Omega(n^{k/(k+1)})$.
\end{prp}

\pfstart
Let $\cert$ be the $k$-subset certificate structure.  Define $\alpha_S(M)$ as
\[
{n\choose k}^{-1/2}\max\sfig{n^{k/(k+1)} - |S|,\; 0}
\]
if $S\notin M$, and as 0 otherwise.  

Let us prove that~\refeqn{alphaOne} holds up to a constant factor.  Take any $S\subset[n]$ and let $j$ be any element not in $S$.  If $|S|\ge n^{k/(k+1)}$, then $\alpha_S(M) = \alpha_{S\cup\{j\}}(M) = 0$ and we are done.  Thus, we further assume $|S|< n^{k/(k+1)}$.  There are ${n\choose k}$ choices of $M$.  If $S\cup\{j\}\notin M$, then the value of $\alpha_S(M)$ changes by ${n\choose k}^{-1/2}$ as the size of $|S|$ increases by 1.  Also, there are at most ${|S|\choose k-1}\le n^{k(k-1)/(k+1)}$ choices of $M\in\cert$ such that $S\notin M$ and $S\cup\{j\}\in M$.  For each of them, the value of $\alpha_S(M)$ changes by at most ${n\choose k}^{-1/2}n^{k/(k+1)}$.  Thus,
\[
\sum_{M\in\cert} (\alpha_S(M) - \alpha_{S\cup\{j\}}(M))^2 \le 
{n\choose k}^{-1} 
\sk[{n\choose k}\cdot1 + n^{k(k-1)/(k+1)}n^{2k/(k+1)} ] 
= O(1).
\]

On the other hand, for the objective value~\refeqn{alphaObjective}, we have
\[
\sqrt{\sum_{M\in\cert} \alphaz(M)^2} = n^{k/(k+1)}.\qedhere
\]
\pfend

Ref.~\cite{lee:learningKdistPrior, zhu:learning} show that the corresponding upper bound is $O(n^{k/(k+1})$, thus the result of \refprp{ksum} is tight.  Moreover, \refthm{boundedly} implies that the complexity of the $k$-sum problem is $\Theta(n^{k/(k+1)})$, a result previously proven in~\cite{spalek:kSumLower}.

\begin{prp}
\label{prp:collision}
The learning graph complexity of the hidden shift (and, hence, the set equality and the collision) certificate structure is $\Omega(n^{1/3})$.
\end{prp}

\pfstart
The proof is similar to the proof of \refprp{ksum}.  Let $\cert$ be the hidden shift certificate structure.  Define $\alpha_M(S)$ as $n^{-1/2}\max\{n^{1/3}-|S|,0\}$ if $S\notin M$, and as 0 otherwise.  Take any $S\subset[n]$, $j\notin S$, and let us prove~\refeqn{alphaOne}.  Again, if $|S|\ge n^{1/3}$, we are done.  Otherwise, there are $n$ choices of $M$ in total, and at most $n^{1/3}$ of them are such that $S\notin M$ and $S\cup\{j\}\in M$.  Thus,
\[
\sum_{M\in\cert} (\alpha_S(M) - \alpha_{S\cup\{j\}}(M))^2 \le 
\frac1n \sk[n\cdot 1 + n^{1/3}n^{2/3}] = O(1).
\]
The objective value~\refeqn{alphaObjective} is $n^{1/3}$.  For the set equality and collision certificate structures, just assign $\alpha_S(M)=0$ for all $M$ that are not in the hidden shift certificate structure.
\pfend

The result of this proposition is also tight.  The corresponding upper bound can be derived by similar methods as used for the $k$-sum problem in Ref.~\cite{lee:learningKdistPrior, zhu:learning}.  We omit the precise construction.

%Thus, we can see that the proofs of these propositions follows closely the best learning graphs and prove they indeed are the best ones.  Also, in the expressions appears the speciality: the ratio of all elements in $\cert$ to the number of ones having particular property ($S\notin M$, but $S\cup\{j\}\in M$).  However, even for the case of the triangle, the complete proof becomes very tedious.

\begin{prp}
\label{prp:triangle}
The learning graph (and, hence, the quantum query) complexity of the triangle certificate structure is $\Omega(n^{9/7}/\sqrt{\log n})$.
\end{prp}

The best known upper bound is $O(n^{9/7})$ as proven in Ref.~\cite{lee:learningTriangle}.  The proof of the lower bound is rather bulky, and essentially proceeds by showing, in a formal way, that all possible strategies of constructing the upper bound fail.

\pfstart[Proof of \refprp{triangle}]
 Let $E=\{uv\mid 1\le u < v\le n\}$ be the set of input variables (potential edges of the graph).
Let $\cert$ be the triangle certificate structure.  We will construct a feasible solution to~\refeqn{learningDual} (with $[n]$ replaced by $E$) in the form
\begin{equation}
\label{eqn:form}
\alpha_S(M) = 
\begin{cases}
\max\{n^{-3/14} - n^{-3/2}|S| - \sum_{i=1}^k g_i(S,M),\; 0\},& S\notin M;\\
0,&\text{otherwise;}
\end{cases}
\end{equation}
where $g_i(S,M)$ is a non-negative function such that $g_i(\emptyset,M)=0$ and $g_i(S,M)\le n^{-3/14}$.  The value of~\refeqn{alphaObjective} is ${\sqrt{{n\choose 3}}}\; n^{-3/14} = \Omega(n^{9/7})$.  The hard part is to show that~\refeqn{alphaOne} holds up to logarithmic factors.  It is easy to see that $\alpha_S(M) = 0$ if $|S|\ge n^{9/7}$, hence, we will further assume $|S|\le n^{9/7}$.

% We first start with {\em partial} solutions of the form~\refeqn{form}.  This means the following.
 For $S\subset E$ and $j\in E\setminus S$, let $F(S,j)$ denote the subset of $M\in\cert$ such that $S\notin M$, but $S\cup\{j\}\in M$.
 We decompose $F(S,j) = F_1(S,j)\sqcup \cdots\sqcup F_k(S,j)$ as follows.
 Each $M\in\cert$ is defined by three vertices $a,b,c$ forming the triangle: $S\in M$ if and only if $ab,ac,bc\in S$.  An input index $j\in E$ satisfies $S\notin M$ and $S\cup\{j\}\in M$ only if $j\in\{ab,ac,bc\}$.  We specify to which of $F_i(S,j)$ an element $M\in F(S,j)$ belongs by the following properties:
\itemstart
\item to which of the three possible edges, $ab$, $ac$ or $bc$, the new edge $j$ is equal, and
\item the range to which the degree in $S$ of the third vertex of the triangle belongs: $[0,n^{3/7}]$, $[n^{3/7}, 2n^{3/7}]$, $[2n^{3/7}, 4n^{3/7}]$, $[4n^{3/7}, 8n^{3/7}]\dots$
\itemend 
Hence, $k\approx 12/7 \log_2 n$.  For notational convenience, let $j=bc$.  Then, the second property is determined by $\deg a = \deg_S a$, the degree of $a$ in the graph with edge set $S$.
 
% these three edges $j$ is equal.  For notational convenience, let it be $bc$.
% We define $F_i(S,j)$ using bounds on the degree $\deg a = \deg_S a$ of the third vertex of the triangle (in our case, $a$) in the graph having $S$ as the edge set: 
% $F_1(S,j)$ consists of all $M\in F(S,j)$  such that $\deg a\le n^{3/7}$ and, for $i\in[2,k]$, $F_i(S,j)$ consists of all $M\in F(S,j)$ such that $2^{i-2}n^{3/7}<\deg a\le 2^{i-1}n^{3/7}$. Hence we choose $k=\lceil 1+4/7\log_2n\rceil$.

For $i\in[k]$, we will define $g_i(S,M)$ so that, for all $S\subset E$ of size at most $n^{9/7}$ and $j\in E\setminus S$:
\begin{equation}
\label{eqn:partial1}
\sum_{M\in\cert\setminus F(S,j)} \sA[g_i(S,M)-g_i(S\cup\{j\}, M)]^2 = O(1)
\end{equation}
and
\begin{equation}
\label{eqn:partial2}
\sum_{M\in F_i(S,j)} \sA[n^{-3/14} - g_i(S,M)]^2 = O(1).
\end{equation}
Let $g_0(S,M)=n^{-3/2}|S|$, for which~\refeqn{partial1} holds.
Even more, we will show that the set $K=K(S,j)$ of $i\in[0,k]$ such that~\refeqn{partial1} is non-zero has size $O(1)$.
 Thus, for the left hand side of~\refeqn{alphaOne}, we will have 
\begin{equation*}
\begin{split}
\sum_{M\in\cert} (\alpha_S(M) - \alpha_{S\cup\{j\}}(M))^2 \le &\;
|K| \sum_{i\in K}\sum_{M\in\cert\setminus F(S,j)} \sA[g_i(S,M) - g_i(S\cup\{j\}, M)]^2 \\
& + \sum_{i=1}^k \sum_{M\in F_i(S,j)} \sA[ n^{-3/14} - g_i(S,M) ]^2,
\end{split}
\end{equation*}
where the former term on the right hand size is $O(1)$ and the latter one is $O(\log n)$.
By scaling all $\alpha_S(M)$ down by a factor of $O(\sqrt{\log n})$, we obtain a feasible solution to~\refeqn{learningDual} with the objective value $\Omega(n^{9/7}/\sqrt{\log n})$.

It remains to construct the functions $g_i(S,M)$.  
In the following, let $\mu(x)$ be the median of $0$, $x$, and $1$, i.e., $\mu(x) = \max\{0,\min\{x,1\}\}$.
The first interval of $\deg a$ will be considered separately from the rest.

% % %

%\paragraph{Case $i=1$}
\paragraph{First interval}
  Assume the condition $\deg a \le n^{3/7}$.  Define
\begin{equation}
\label{eqn:g1}
g_i(S, M) = \begin{cases}
n^{-3/14}\,\mu(2 - n^{-3/7}\deg a),&\text{$ab,ac\in S$;}\\
0,&\text{otherwise.}
\end{cases}
\end{equation}
Clearly, $g_i(\emptyset,M)=0$ and $g_i(S,M)\ge 0$.  There are two cases how $g_i(S,M)$ may be influenced.  We show that the total contribution to~\refeqn{partial1} is $O(1)$.
\itemstart
\item It may happen if $|\{ab,ac\}\cap S| = 1$ and $j\in \{ab,ac\}$, i.e., the transition from the second case of~\refeqn{g1} to the first one happens.  Moreover, $g_1(S,M)$ changes only if $\deg a\le 2n^{3/7}$.  Then $j$ identifies two vertices of the triangle, and the third one is among the neighbours of an endpoint of $j$ having degree at most $2n^{3/7}$.  Thus, the total number of $M$ satisfying this scenario is at most $4n^{3/7}$.  The contribution to~\refeqn{partial1} is at most $O(n^{3/7})(n^{-3/14})^2 =O(1)$.
\item Another possibility is that $ab,ac\in S$ and $\deg a$ changes.  In this case, $a$ is determined as an endpoint of $j$, and $b$ and $c$ are among its at most $2n^{3/7}$ neighbours.  The number of $M$ influenced is $O(n^{6/7})$, and the contribution is $O(n^{6/7})(n^{-9/14})^2 = o(1)$.
\itemend
Finally, we have to show that~\refeqn{partial2} holds.  If $M$ satisfies the condition, then $ab,ac\in S$ and $\deg a\le n^{3/7}$.  In this case, the left hand side of~\refeqn{partial2} is 0.

%\paragraph{Case $i\in[2,k]$}
\paragraph{Other intervals}
  Now assume the condition $d < \deg a\le 2d$ with $d\ge n^{3/7}$. Define a piece-wise linear function $\tau$ as follows
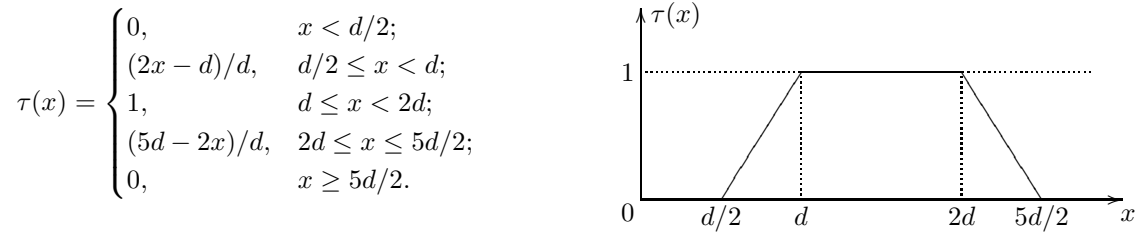
\[
\tau(x) =
\begin{cases}
0,& x<d/2;\\
(2x-d)/d,& d/2\le x<d;\\
1,& d\le x < 2d;\\
(5d-2x)/d,& 2d\le x\le 5d/2;\\
0,& x\ge 5d/2.
\end{cases}
\xygraph{!{0;<1pc,0pc>:}
[d(3)r(5)](
	[ld(.4)]*{0},
	:[u(6)][d(.2)r(1.1)]*{\tau(x)},
	:[r(15)][r(.2)d(.5)]*{x},
	[u(4)](
		[l(.4)]*{1},
		-@{.}[r(14)]
		),
	[d(.8)][r(2.5)]*{\smash{d/2}}[r(2.5)]*{\smash{d}}[r(5)]*{\smash{2d}}[r(2.5)]*{\smash{5d/2}},
	[r(2.5)]-[r(2.5)u(4)](-@{.}[d(4)])-[r(5)](-@{.}[d(4)])-[r(2.5)d(4)]
	)
}
\]
It can be interpreted as a continuous version of the indicator function that a vertex has a right degree.  Define
\[
\nu(S,M) = \sum_{v\in N(b)\cap N(c)} \tau(\deg v),
\]
where the sum is over the common neighbours of $b$ and $c$.  Let
\[
g_i(S,M) = n^{-3/14}\,\mu\sB[\min\sfig{\frac{2\deg a}{d}, \frac{\nu(S,M)}{n^{3/7}} }-1].
\]
Let us consider how $g_i(S, M)$ may change and how this contributes to~\refeqn{alphaOne}.
Now there are three cases how $g_i(S,M)$ may be influenced.  We again show that the total contribution to~\refeqn{partial1} is $O(1)$.

\itemstart
\item It may happen that $j$ is incident to a common neighbour of $b$ and $c$, and thus $\nu(S)$ may change.  This means $b$ and $c$ are among the neighbours of an endpoint of $j$ of degree at most $5d/2$.  Hence, this affects $O(nd^2)$ different $M$.  The contribution is $O(nd^2)(n^{-9/14}/d)^2 = o(1)$.

\item The set $N(b)\cap N(c)$ may increase.  This causes a change in $g_i(S, M)$ only under the following circumstances.  The new edge $j$ is incident to $b$ or $c$.  The second vertex in $\{b,c\}$ is among $\Theta(d)$ neighbours of the second end-point of $j$.  Finally, $\deg a\ge d/2$, that together with $|S|\le n^{9/7}$ implies that there are $O(n^{9/7}/d)$ choices for $a$.  Altogether, the number of $M$ affected by this is $O(n^{9/7})$, and the change in $g_i(S, M)$ does not exceed $n^{-9/14}$.  The contribution is $O(1)$.

\item The degree of $a$ may change.  Let us calculate the number $P$ of possible pairs $b$ and $c$ affected by this.  There is a change in $g_i(S, M)$ only if $b$ and $c$ are connected to at least $n^{3/7}$ vertices of degrees between $d/2$ and $5d/2$.  Denote the set of these vertices by $A$.  Since $|S|\le n^{9/7}$, we have $|A|=O(n^{9/7}/d)$.  

Let us calculate the number of paths of length 2 in $S$ having the middle vertex in $A$.   On one hand,  this number is at least $Pn^{3/7}$.  On the other hand, it is at most $O(d^2 |A|) = O(dn^{9/7})$.  Thus, $P = O(dn^{6/7})$.  Since $a$ is determined as an end-point of $j$, the contribution is $O(dn^{6/7})(n^{-3/14}/d)^2 = O(1)$, as $d\ge n^{3/7}$.
\itemend

Finally, $j$ may be the last edge of the triangle.  We know that $\deg a> d$, hence, either $n^{-3/14} - g_i(S,M) = 0$, or $\nu(S,M)\le 2n^{3/7}$, in which case, there are $O(n^{3/7})$ choices of $a$ satisfying the condition.  Hence, the left hand side of~\refeqn{partial2} is $O(n^{3/7}) (n^{-3/14})^2 = O(1)$.

%A careful analysis shows that there is a subset $K$ of conditions with $|K|=O(1)$ and $g_i(S,M) - g_i(S\cup\{j\}, M)= 0$ for all $i\notin K$ and $M\in\cert\setminus F(S,j)$.  (In two words: in all three subcases, the value of $d$ may be determined with some ambiguity from $j$.) 
If $g_i(S,M) - g_i(S\cup\{j\}, M)\ne 0$, then, in the first three cases, the value of $d$, up to a small ambiguity, may be determined from the degree of one of the end-points of $j$.  Hence, the set $K=K(S,j)$, as stated previously in the proof, exists.
\pfend

Automatically, we obtain that the quantum query complexity of the {\em triangle sum} problem is $\widetilde{\Omega}(n^{9/7})$.  Thus, any quantum query algorithm, willing to improve the $O(n^{9/7})$ bound for the triangle detection problem, will have to take differences between the triangle detection and triangle sum problems into consideration.

\section{Lower Bound}
\label{sec:lower}
In this section, we prove Theorems~\ref{thm:certificates} and~\ref{thm:boundedly}.  The results are strongly connected: In the second one we prove a stronger statement from stronger premisses.  As a consequence, the proofs also have many common elements.

This section is organized as follows.  In \refsec{adversary}, we recall the adversary method that we use to prove the lower bound.  In the proofs, we will define a number of matrices and argue about their spectral properties.  For convenience, we describe the main parameters of the matrices, such as the labelling of their rows and columns, as well as their mutual relationships in one place, \refsec{outline}.  In \refsec{common}, we state the intermediate results important to both Theorems~\ref{thm:certificates} and~\ref{thm:boundedly}.  In~\refsec{boundedly}, we finish the proof of \refthm{boundedly}.  In \refsec{fourier}, we recall the definition and main properties of the Fourier basis, and define the important notion of the Fourier bias.  Finally, in \refsec{general}, we prove \refthm{certificates}.

\subsection{Adversary Bound}
\label{sec:adversary}
The adversary method is one of the main techniques for proving lower bounds on quantum query complexity.  First developed by Ambainis~\cite{ambainis:adv}, it was later strengthened by H\o yer \etal~\cite{hoyer:advNegative}. After that, the adversary bound was proven to be optimal by Reichardt {\em et al.}~\cite{reichardt:advTight, lee:stateConversion}.  In this paper, we use a variation of the adversary bound from Ref.~\cite{spalek:kSumLower}.

\mycommand{domain}{\widetilde{\cD}}

\begin{defn}
\label{defn:adversary}
Let $f$ be a function $f\colon \cD\to \{0,1\}$ with domain $\cD\subseteq [q]^n$.  Let $\domain$ be a set of pairs $(x,a)$ with the property that the first element of each pair belongs to $\cD$, and $\domain_i = \{(x,a)\in \domain \mid f(x)=i\}$ for $i\in\{0,1\}$.  An {\em adversary matrix} for the function $f$ is a non-zero real $\domain_1\times\domain_0$ matrix $\Gamma$. And, for $j\in[n]$, let $\Delta_j$ denote the $\domain_1\times \domain_0$ matrix defined by
\[ \Delta_j\elem[(x,a),(y,b)] = \begin{cases} 0,& x_j=y_j; \\ 1,&\text{otherwise}. \end{cases} \]
\end{defn}

\begin{thm}[Adversary bound \cite{hoyer:advNegative, spalek:kSumLower}]
\label{thm:adv}
In the notations of Definition~\ref{defn:adversary}, the quantum query complexity of $f$ is $\Omega(\Adv(f))$, where
\begin{equation}\label{eqn:adversary}
\Adv(f) = \sup_{\Gamma} \frac{\norm|\Gamma|}{\max_{j\in n} \norm|\Gamma\circ\Delta_j| }
\end{equation}
with the maximization over all adversary matrices for $f$, and $\norm|\cdot|$ is the spectral norm.
\end{thm}

The following result is very useful when proving lower bounds using the adversary method	.
\begin{lem}[\cite{lee:stateConversion}]
\label{lem:gamma}
Let $\Delta_j$ be as in \refdefn{adversary}.  Then, for any matrix $A$ of the same size, 
\[
\norm|A\circ \Delta_j| \le 2 \norm|A|.
\]
\end{lem}
\mycommand{dar}{\stackrel{\Delta_j}{\longmapsto}}

We will use it to replace $\Gamma\circ \Delta_j$ in the denominator of~\refeqn{adversary} with a matrix $\Gamma'$ such that $\Gamma\circ\Delta_j = \Gamma'\circ\Delta_j$.  By \reflem{gamma}, this gives the same result up to a factor of 2.  We will denote this relation between matrices by $\Gamma\dar\Gamma'$.

\subsection{Outline}
\label{sec:outline}
\begin{figure}[tbh]
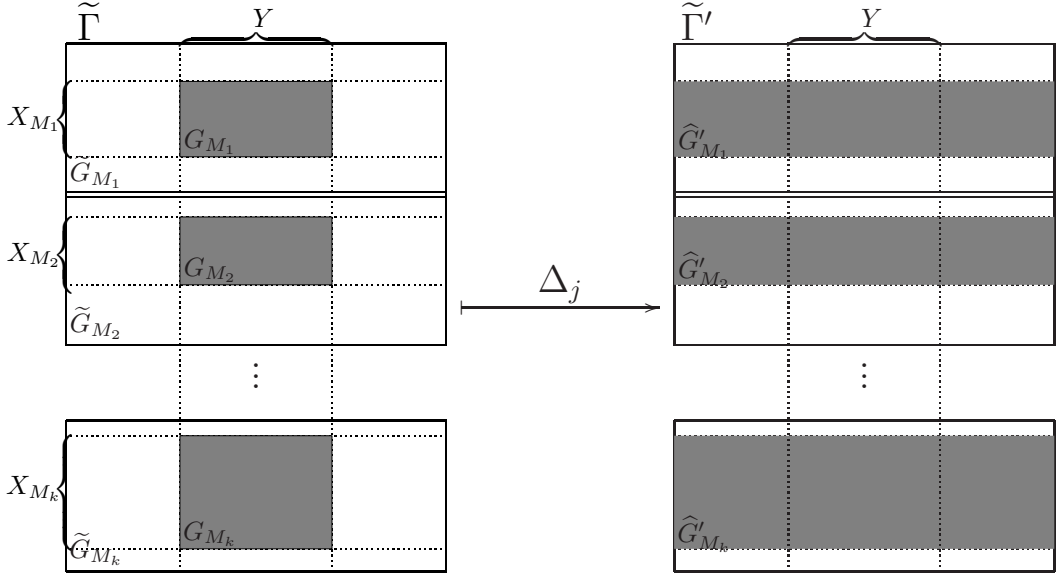

\release{
\[
\xy %<0.12cm,0cm>:
="st"
 @={+(0,0), -(0,19.7),-(0,0.6), -(0,19.7), -(0,10), -(0,20)},
 @@{; p +(50,0) **@{-}},
 "st" 
 @={+(0,0), +(50,0)},
 @@{; p -(0,40) **@{-}, -(0,10); p-(0,20) **@{-}},
 "st" 
 @={+(15,0)="d", +(20,0)="d2"},
 @@{; p -(0,70) **@{.}},
 "st"+(25,0.4)."d" ; "d2" **\frm{^\}}; +(1,3) *{Y},
 "st" 
 @={-(0,5)="a", -(0,10)="a2", "st"-(0,23)="b", -(0,9)="b2", "st"-(0,52)="c", -(0,15)="c2"},
 @@{; p +(50,0) **@{.}},
 "st"+(3,3) *\txt{\Large$\tGamma$},
 "a"-(0.2,5)."a" ; "a2" **\frm{\{} ; -(4,0) *{X_{M_1}},
 "b"-(0.2,5)."b" ; "b2" **\frm{\{} ; -(4,0) *{X_{M_2}},
 "c"-(0.2,7)."c" ; "c2" **\frm{\{} ; -(4,0) *{X_{M_k}},
 "a"+(15,0); "a2"+(35,0).p *[F*:gray]\frm{-},
 "b"+(15,0); "b2"+(35,0).p *[F*:gray]\frm{-},
 "c"+(15,0); "c2"+(35,0).p *[F*:gray]\frm{-},
 "st"+(4,-17) *{\tG_{M_1}}, -(0,20) *{\tG_{M_2}}, -(0,30) *{\tG_{M_k}},
 "a2"+(19,2) *{G_{M_1}}, "b2"+(19,2) *{G_{M_2}}, "c2"+(19,2) *{G_{M_k}},
 "st"+(25,-43) *\txt{\Large$\vdots$},
"st"+(80,0)="st",
 @={+(0,0), -(0,19.7),-(0,0.6), -(0,19.7), -(0,10), -(0,20)},
 @@{; p +(50,0) **@{-}},
 "st" 
 @={+(0,0), +(50,0)},
 @@{; p -(0,40) **@{-}, -(0,10); p-(0,20) **@{-}},
 "st" 
 @={-(0,5)="a", -(0,10)="a2", "st"-(0,23)="b", -(0,9)="b2", "st"-(0,52)="c", -(0,15)="c2"},
 @@{; p +(50,0) **@{.}},
 "st"+(3,3) *\txt{\Large$\tGamma'$},
% "a"-(0.2,5)."a" ; "a2" **\frm{\{} ; -(4,0) *{X_{M_1}},
% "b"-(0.2,5)."b" ; "b2" **\frm{\{} ; -(4,0) *{X_{M_2}},
% "c"-(0.2,7)."c" ; "c2" **\frm{\{} ; -(4,0) *{X_{M_\ell}},
 "a2"+(50,0)."a" *[F*:gray]\frm{},
 "b2"+(50,0)."b" *[F*:gray]\frm{},
 "c2"+(50,0)."c" *[F*:gray]\frm{},
 "st" 
 @={+(15,0)="d", +(20,0)="d2"},
 @@{; p -(0,70) **@{.}},
 "st"+(25,0.4)."d" ; "d2" **\frm{^\}}; +(1,3) *{Y},
 "a2"+(4,2) *{\hG_{M_1}}, "b2"+(4,2) *{\hG_{M_2}}, "c2"+(4,2) *{\hG_{M_k}},
 "st"+(25,-43) *\txt{\Large$\vdots$},
"st"-(28,35); p+(13,3)*\txt{\Large$\Delta_j$} ;
\ar @{|->} +(26,0)
\endxy
\]

\caption{The relationships between matrices used in \refsec{lower}.  The parts marked in grey form the matrix $\Gamma$ on the left, and $\hGamma$ on the right.  Note that they are {\em not} submatrices of $\tGamma$ and $\tGamma'$, respectively: 
They have additional multiplicative factor as specified in~\refeqn{GammaElements} and~\refeqn{hGM}.
}}
\label{fig:matrices}
\end{figure}

Let us briefly outline how Theorems~\ref{thm:certificates} and~\ref{thm:boundedly} are proven.  Let $\cert$ denote the certificate structure.  Let $\alpha_S(M)$ satisfy~\refeqn{learningDual}, and be such that~\refeqn{alphaObjective} equals the learning graph complexity of $\cert$.  We define an explicit function $f\colon\cD\to \{0,1\}$ with $\cD\subseteq [q]^n$ having the objective value~\refeqn{alphaObjective} of program~\refeqn{learningDual} as a lower bound on its quantum query complexity.  The latter is proven using the adversary bound, \refthm{adv}.  For that, we define a number of matrices, as illustrated in \reffig{matrices}.
%And the adversary matrix $\Gamma$ is constructed from the values of $\alpha_S(M)$.

\mycommand{tGamma}{\widetilde{\Gamma}}
\mycommand{tG}{\widetilde{G}}
\mycommand{hG}{\widehat{G}'}
\mycommand{hGamma}{\widehat{\Gamma}'}

\paragraph{Matrix $\tGamma$} 
At first, we construct a matrix $\tGamma$ satisfying the following properties.  Firstly, it has rows labelled by the elements of $[q]^n\times\cert$, and columns labelled by the elements of $[q]^n$.  Thus, if we denote $\cert = \{\marked_1,\dots,\marked_k\}$, the matrix $\tGamma$ has the following form
\begin{equation}
\label{eqn:tGamma}
\tGamma = 
\begin{pmatrix}
\tG_{\marked_1}\\ \tG_{\marked_2}\\\vdots\\ \tG_{\marked_{k}}
\end{pmatrix},
\end{equation}
where each $\tG_\marked$ is an $[q]^n\times[q]^n$-matrix.  Next, $\|\tGamma\|$ is at least the objective value~\refeqn{alphaObjective}.  And finally, for each $j\in[n]$, there exists $\tGamma'$ such that $\tGamma\dar\tGamma'$ and $\|\tGamma'\|\le 1$.  
%Here, $\dar$ is as described after \reflem{gamma}.  
The matrix $\tGamma'$ has a decomposition into blocks $\tG'_M$ similar to~\refeqn{tGamma}.

Thus, $\tGamma$ has a good value of~\refeqn{adversary}.  But, we cannot use it, because it is not an adversary matrix: It uses all possible inputs as labels of both rows and columns.  However, due to the specific way $\tGamma$ is constructed, we will be able to transform $\tGamma$ into a true adversary matrix $\Gamma$ such that the value of~\refeqn{adversary} is still good.  Before we describe how we do it, let us outline the definition of the function $f$.

\mycommand{sz}{\ell}
\mycommand{fii}{f^{-1}}
\paragraph{Defining the function}
Let $M$ be an element of the certificate structure $\cert$.  Let $A_{M}^{(1)},\dots,A_{M}^{(\sz(M))}$ be all the inclusion-wise minimal elements of $M$.  (In a boundedly generated certificate structure, $M$ has only one inclusion-wise minimal element $\certM$.)  For each $A_{M}^{(i)}$, we choose an orthogonal array $T_{M}^{(i)}$ of length $|A_{M}^{(i)}|$ over the alphabet $[q]$, and define
\begin{equation}
\label{eqn:Xm}
X_M = \sfig{x\in[q]^n\mid \mbox{$x_{A_{M}^{(i)}}\in T_{M}^{(i)}$ for all $i\in[\sz(M)]$} }.
\end{equation}
The orthogonal arrays are chosen so that $X_M$ is non-empty and satisfies the following {\em orthogonality property}:
\begin{equation}
\label{eqn:orthogonality}
\forall S\in 2^{[n]}\setminus M\;\; \forall z\in [q]^S\;:\; \abs|\strut\{x\in X_M \mid x_S = z \}| = |X_M|/q^{|S|}.
\end{equation}
For boundedly generated certificate structures, this property is satisfied automatically.

The set of positive inputs is defined by $\fii(1) = \bigcup_{M\in\cert} X_M$.  The set of negative inputs is defined by
\begin{equation}
\label{eqn:fii}
\fii(0) = \sfig{x\in[q]^n\mid \mbox{$x_{A_{M}^{(i)}}\notin T_{M}^{(i)}$ for all $M\in\cert$ and $i\in[\sz(M)]$} }.
\end{equation}
It is easy to see that $f$ has $\cert$ as its certificate structure.  The parameters will be chosen so that $|\fii(0)|=\Omega(q^n)$.  %If $\cert$ is boundedly generated, the function $f$ is total.

\paragraph{Remaining matrices}
Let us define $X = \{(x,M)\in[q]^n\times\cert \mid x\in X_M\}$ and $Y = \fii(0)$.  The matrix $\Gamma$ is an $X\times Y$ matrix defined by
\begin{equation}
\label{eqn:GammaElements}
\Gamma\elem[(x,M), y] = \sqrt{\frac{q^n}{|X_M|}}\; \tGamma\elem[(x,M),y].
\end{equation}
Thus, $\Gamma$ consists of blocks $G_M$, like in~\refeqn{tGamma}, where $G_M = \sqrt{q^n/|X_M|}\; \tG_M\elem[X_M,Y]$.  (The latter notation stands for the submatrix formed by the specified rows and columns).  We also show that $\|\Gamma\|$ is not much smaller than $\|\tGamma\|$.

The matrix $\Gamma'$ is obtained similarly from $\tGamma'$.  It is clear that $\tGamma\dar\tGamma'$ implies $\Gamma\dar\Gamma'$.  We show that the norm of $\Gamma'$ is small by showing that $\|\hGamma\| = O(\|\tGamma'\|)$ where $\hGamma$ is an $X\times [q]^n$-matrix with 
\[
\hGamma\elem[(x,M),y] = \sqrt{\frac{q^n}{|X_M|}}\; \tGamma'\elem[(x,M),y].
\]
As $\Gamma'$ is a submatrix of $\hGamma$ and $\|\tGamma'\|\le 1$, we obtain that $\|\Gamma'\| = O(1)$ as required.  We denote the blocks of $\hGamma$ by $\hG_M$.  That is, 
\begin{equation}
\label{eqn:hGM}
\hG_M = \sqrt{\frac{q^n}{|X_M|}}\; \tG_M'\elem[X_M,{[q]^n}].
\end{equation}

\subsection{Common Parts of the Proofs}
\label{sec:common}

%We apply the adversary bound~\refthm{adv}.  The construction is similar to the one in~\cite{spalek:kSumLower}. 
Let $e_0,\dots,e_{q-1}$ be an orthonormal basis of $\C^q$ such that $e_0=1/\sqrt{q}(1,\dots,1)$.  Denote $E_0 = e_0e_0^*$ and $E_1 = \sum_{i>0} e_ie_i^*$.  These are $q\times q$ matrices. All entries of $E_0$ are equal to $1/q$, and the entries of $E_1$ are given by
\begin{equation}
\label{eqn:E1entries}
E_1\elem[x,y] = \begin{cases}
1-1/q,& x=y;\\
-1/q,& x\ne y.
\end{cases}
\end{equation}
For a subset $S\subseteq[n]$, let $E_S$ denote $\bigotimes_{j\in[n]} E_{s_j}$ where $s_j=1$ if $j\in S$, and $s_j=0$ otherwise.  These matrices are orthogonal projectors:
\begin{equation}
\label{eqn:ESOrthogonal}
E_SE_{S'} =
\begin{cases}
E_S,& S=S'\\
0,&\text{otherwise.}
\end{cases}
\end{equation}

We define the matrices $\tG_M$ from~\refeqn{tGamma} by
\begin{equation}
\label{eqn:tG}
\tG_\marked = \sum_{S\subseteq[n]} \alpha_S(\marked) E_S,
\end{equation}
where $\alpha_S(M)$ are as in~\refeqn{learningDual}. 

\begin{lem}
\label{lem:GammaNorm}
If $\tGamma$ and $\Gamma$ are defined as in \refsec{outline}, all $X_M$ satisfy the orthogonality property~\refeqn{orthogonality} and $|Y|=\Omega(q^n)$, then 
\begin{equation}
\label{eqn:GammaNorm}
\|\Gamma\| = \Omega\sD[\sqrt{\sum_{M\in\cert} \alphaz(M)^2}].
\end{equation}
\end{lem}

\mycommand{summa}{\mathrm{s}}
\mycommand{gm}{G_\marked}
\mycommand{tgm}{\tG_\marked}
\pfstart
Recall that $\gm = \sqrt{q^n/|X_M|}\tgm\elem[X_M,Y]$, hence, by~\refeqn{tG}:
\[%begin{equation}
%\label{eqn:tG2}
\gm = \sqrt{\frac{q^n}{|X_M|}}\; \alphaz(M) E_0^{\otimes n}\elem[X_M,Y] + \sqrt{\frac{q^n}{|X_M|}} 
\sum_{S\ne\emptyset} \alpha_S(\marked) E_S\elem[X_M,Y].
\]%end{equation}
Let us calculate the sum $\summa(G_M)$ of the entries of $G_M$.
In the first term, each entry of $E_0^{\otimes n}$ equals $q^{-n}$.  There are $|X_M|$ rows and $|Y|$ columns in the matrix, hence, the sum of the entries of the first term is $\sqrt{|X_M|/q^n}\;|Y|\alphaz(M)$.  

We claim that, in the second term, $\summa\s[\strut \alpha_S(M){E_S\elem[X_M,Y]}]=0$ for all $S\ne \emptyset$.  Indeed, if $S\in M$, then $\alpha_S(M)=0$ by~\refeqn{alphaZero}.  Otherwise,
\[
\summa(E_S\elem[X_M,Y]) = \sum_{y\in Y}\sum_{x\in X_M} E_S\elem[x, y] = 
q^{|S|-n} \sum_{y\in Y}\sum_{x\in X_M} E_1^{\otimes|S|} \elem[x_S, y_S] = 
\frac{|X_M|}{q^{n}} \sum_{y\in Y} \sum_{z\in [q]^S} E_1^{\otimes|S|} \elem[z, y_S] = 0.
\]
%\[
%\s(E_S\elem[X_M,Y]) = \sum_{y\in Y} \s(E_S\elem[X_M,\{y\}]) = \sum_{y\in Y} \frac{|X_M|}{q_S} \s(E_1^{\otimes|S|}\elem[{[q]}^S, \{{y\elem[S]}\}] ) = 0.
%\]
(On the third step, the orthogonality condition~\refeqn{orthogonality} is used.  On the last step, we use that the sum of the entries of every column of $E_1^{\otimes k}$ is zero if $k>0$.)  Summing up,
\[
\summa(\gm) = \sqrt{\frac{|X_M|}{q^n}}\;|Y|\alphaz(M).
\]

We are now ready to estimate $\norm|\Gamma|$.  Define two unit vectors $u\in\R^X$ and $v\in\R^Y$ by 
\[
u\elem[(x,M)] = \frac{\alphaz(M)}{\sqrt{|X_M|\sum_{M\in\cert} \alphaz(M)^2}}\qquad\text{and}\qquad
v\elem[y] = \frac1{\sqrt{|Y|}}
\]
for all $(x,M)\in X$ and $y\in Y$.  Then,
%\begin{equation}
%\label{eqn:GammaNorm}
\[
\norm|\Gamma| \ge u^*\Gamma v = \frac{\sum_{M\in\cert} \alphaz(M)\summa(G_M)}{\sqrt{|X_M|\;|Y|\sum_{M\in\cert} \alphaz(M)^2}} = \sqrt{\frac{|Y|}{q^n}\sum_{M\in\cert} \alphaz(M)^2} = \Omega\sD[\sqrt{\sum_{M\in\cert} \alphaz(M)^2}] .\qedhere
\]%end{equation}
%by the assumption on the size of $|Y|$.
\pfend

In the remaining part of this section, we define the transformation $\tGamma\dar\tGamma'$ and state some of the properties of $\tGamma'$ that will be used in the subsequent sections.
Using~\refeqn{E1entries}, we can define the action of $\Delta$ on $E_0$ and $E_1$ by 
\[%begin{equation}
%\label{eqn:delta}
E_0\stackrel{\Delta}{\longmapsto} E_0\qquad\text{and}\qquad E_1\stackrel{\Delta}{\longmapsto} -E_0.
\]%end{equation}
We define $\tGamma'$ by applying this transformation to $E_0$ and $E_1$ in the $j$th position in the tensor product of~\refeqn{tG}.  The result is again a matrix of the form~\refeqn{tGamma}, but with each $\tgm$ replaced by
\begin{equation}
\label{eqn:gmPrim}
\tgm' = \sum_{S\subseteq[n]} \beta_S(M)E_S,
\end{equation}
where $\beta_S(M) = \alpha_S(M) - \alpha_{S\cup\{j\}}(M)$.  In particular, $\beta_S(M)=0$ if $j\in S$ or $S\in M$.  Thus,
\begin{equation}
\label{eqn:GammaPrim}
(\tGamma')^*\tGamma' = \sum_{M\in\cert} (\tgm')^*\tgm' = \sum_{S\in 2^{[n]}} \sB[\sum_{M\in\cert} \beta_S(M)^2] E_S.
\end{equation}
In particular, we obtain from~\refeqn{alphaOne} that $\|\tGamma'\|\le 1$.

\subsection{Boundedly generated certificate structures}
\label{sec:boundedly}
In this section, we finish the proof of \refthm{boundedly}.  In the settings of the theorem, the orthogonal arrays $T_M^{(i)}$ in~\refeqn{Xm} are already specified.  Since each $M\in\cert$ has only one inclusion-wise minimal element $\certM$, we drop all upper indices $(i)$ in this section.  

From the statement of the theorem, we have $|X_M| = q^{n-1}$, in particular, they are non-empty.  Also, $X_M$ satisfy the orthogonality property~\refeqn{orthogonality}, and, by~\refeqn{fii}, we have
\begin{equation}
\label{eqn:Ysize}
|Y| = \absC|[q]^n\setminus\bigcup_{M\in\cert} X_M| \ge q^n - \sum_{M\in\cert} |X_M| = q^n - |\cert|q^{n-1} \ge \frac{q^n}2.
\end{equation}
Thus, the conditions of \reflem{GammaNorm} are satisfied, and~\refeqn{GammaNorm} holds.

Recall from \refsec{outline} that in order to estimate $\|\Gamma'\|$ we consider the matrix $\hGamma$.  The matrix $\Gamma'$ is a submatrix of $\hGamma$, hence, it suffices to estimate $\|\hGamma\|$.  Let $k=\max_{M\in\cert} |\certM|$.  By \refdefn{boundedly}, $k=O(1)$.  

Fix an arbitrary order of the elements in each $\certM = \{a_{M,1},\dots,a_{M,|\certM|}\}$, and let $L_{M,i}$, where $M\in\cert$ and $i\in[k]$, be subsets of $2^{[n]}$ satisfying the following properties:
\itemstart
\item for each $M$, the set $2^{[n]}\setminus M$ is the disjoint union $L_{M,1}\sqcup\cdots\sqcup L_{M,k}$;
\item for each $M$ and each $i\le |\certM|$, all elements of $L_{M,i}$ omit $a_{M,i}$;
\item for each $M$ and each $i$ such that $|\certM|<i\le k$, the set $L_{M,i}$ is empty.
\itemend
Recall that, if $S\subseteq[n]$ and $(s_j)$ is the corresponding characteristic vector, $E_S = \bigotimes_{j\in[n]} E_{s_j}$.  The main idea behind defining $L_{M,i}$s is as follows.

\begin{clm}
\label{clm:LMi}
If $S,S'\in L_{M,i}$, then
\[
(E_S\elem[X_M,{[q]^n}])^* (E_{S'}\elem[X_M,{[q]^n}]) = 
\begin{cases}
E_S/q,& S=S';\\
0,&\text{\rm otherwise.}
\end{cases}
\]
\end{clm}

\pfstart
If we strike out the $a_{M,i}$th element in all elements of $X_M$, we obtain $[q]^{n-1}$ by the definition of an orthogonal array.  All elements of $L_{M,i}$ omit $a_{M,i}$, hence, $E_S$ has $E_0$ in the $a_{M,i}$th position for all $S\in L_{M,i}$.  Thus, the $a_{M,i}$th entries of $x$ and $y$ has no impact on the value of $E_S\elem[x,y]$.  

Let $(s_j)$ and $(s'_j)$ be the characteristic vectors of $S$ and $S'$.  Then,
\[
E_S\elem[X_M,{[q]^n}] = \sC[\bigotimes_{j\in[n]\setminus\{a_{M,i}\}} E_{s_j}]\otimes \frac{e_0^*}{\sqrt{q}}.
\]
(Here $e_0^*$ is on the $a_{M,i}$th element of $[q]^n$.)  Similarly for $S'$, and the claim follows from~\refeqn{ESOrthogonal}.
\pfend

For each $M$, decompose $\tgm'$ from~\refeqn{gmPrim} into $\sum_{i\in[k]} \tG'_{M,i}$, where
\[
\tG'_{M,i} = \sum_{S\in L_{M,i}} \beta_S(M) E_S.
\]
Define similarly to \refsec{outline}, 
\[
\hG_{M,i} = \sqrt{\frac{q^n}{|X_M|}}\; \tG'_{M,i}\elem[X_M,{[q]^n}] = \sqrt{q}
\sum_{S\in L_{M,i}} \beta_S(M) E_S\elem[X_M,{[q]^n}],
\]
and let $\hGamma_i$ be the matrix consisting of $\hG_{M,i}$, for all $M\in\cert$, stacked one on another like in~\refeqn{tGamma}.  Then, $\hGamma = \sum_{i\in [k]} \hGamma_i$.
We have
\[%begin{equation}
%\label{eqn:urav1}
(\hGamma_i)^*\hGamma_i = \sum_{M\in \cert} (\hG_{M,i})^* \hG_{M,i} = \sum_{M\in\cert} \sum_{S\in L_{M,i}} \beta_S(M)^2E_S,
\]%end{equation}
by \refclm{LMi}.  Similarly to~\refeqn{GammaPrim}, we get $\|\hGamma_i\|\le 1$.  By the triangle inequality, $\normS|\hGamma|\le k$, hence, $\norm|\Gamma'| \le k=O(1)$.  Combining this with~\refeqn{GammaNorm}, and using \refthm{adv}, we obtain the necessary lower bound.  This finishes the proof of \refthm{boundedly}.

\subsection{Fourier Basis}
\label{sec:fourier}
\mycommand{group}{Z}
\mycommand{fb}{\chi}
\def \fnorm|#1|{\|#1\|_{\mathrm{u}}}
In \refsec{common}, we defined $e_i$ as an arbitrary orthonormal basis satisfying the requirement that $e_0$ has all its entries equal to $1/\sqrt{q}$.  In the next section, we will specify a concrete choice for $e_i$.  Its construction is based on the Fourier basis we briefly review in this section.

Let $p$ be a positive integer, and $\Z_p$ be the cyclic group of order $p$, formed by the integers modulo $p$. Consider the complex vector space $\C^{\Z_p}$.  The vectors $(\fb_a)_{a\in \Z_p}$, defined by $\fb_a\elem[b] = \ee^{2\pi\ii ab/p}/\sqrt{p}$, form its orthonormal basis.  Note that the value of $\fb_a\elem[b]$ is well-defined because $\ee^{2\pi\ii}=1$.

If $U\subseteq\Z_p$, then the {\em Fourier bias}~\cite{tao:additive} of $U$ is defined by
\begin{equation}
\label{eqn:fnorm}
\fnorm|U|  % = \abs|\max_{a\in\Z_p\setminus\{0\}} \summa(\fb_a\elem[U]) |
= \frac{1}{p}\; \absC| \max_{a\in\Z_p\setminus\{0\}} \sum_{u\in U} \ee^{2\pi\ii au/p} |.
\end{equation}
It is a real number between 0 and $|U|/p$.  In the next section, we will need the following result stating the existence of sets with small Fourier bias and arbitrary density.

\begin{thm}
\label{thm:FourierBias}
For any real $0<\delta<1$, it is possible to construct $U\subseteq \Z_q$ such that $|U|\sim\delta q$, $\fnorm|U| = O(\polylog(q)/\sqrt{q})$ and $q$ is arbitrary large.  In particular, $\fnorm|U| = o(1)$.
\end{thm}

For instance, one may prove a random subset satisfies these properties with high probability~\cite[Lemma~4.16]{tao:additive}.  There also exist explicit constructions~\cite{gillespie:fourierBias}.

\subsection{General Certificate Structures}
\label{sec:general}
In this section, we finish the proof of \refthm{certificates}.  There are two main reasons why it is not possible to prove a general result like~\refthm{boundedly} for arbitrary certificate structures.

A first counterexample is given by \refprp{collision} stating that the learning graph complexity of the hidden shift certificate structure is $\Omega(n^{1/3})$ and the statement at the end of \refsec{examples} that the quantum query complexity of the hidden shift problem is $O(\log n)$.  The proof in \refsec{boundedly} cannot be applied here, because $k$ in the decomposition of $\tG'_M$ into $\sum_{i\in[k]} \tG'_{M,i}$ would not be bounded by a constant.  We solve this by considering much ``thicker'' orthogonal arrays $T_M^{(i)}$.

Next, the orthogonality property~\refeqn{orthogonality} is not satisfied automatically for general certificate structures.  For instance, assume $A_{M}^{(1)} = \{1,2\}$, $A_{M}^{(2)}=\{2,3\}$, and the orthogonal arrays are given by the conditions $x_1=x_2$ and $x_2=x_3$, respectively.  Then, for any input $x$ satisfying both conditions, we have $x_1 = x_3$, and the orthogonality condition fails for $S = \{1,3\}$.

The problem in the last example is that the orthogonal arrays are not independent because $A_{M}^{(1)}$ and $A^{(2)}_{M}$ intersect.  We cannot avoid that $A_{M}^{(i)}$s intersect, but we still can have $T_M^{(i)}$s independent by defining them on independent parts of the input alphabet.  

More formally, let $\sz = \max_{M\in\cert} \sz(M)$, where $\sz(M)$ is defined in \refsec{outline} as the number of inclusion-wise minimal elements of $M$.  We define the input alphabet as $\group = \Z_p^\sz$ for some $p$ to be defined later.  Hence, the size of the alphabet is $q = p^\sz$.  

Let $Q_M^{(i)}$ be an orthogonal array of length $|A_M^{(i)}|$ over the alphabet $\Z_p$.  We will specify a concrete choice in a moment.  From $Q_M^{(i)}$, we define $T_M^{(i)}$ in~\refeqn{Xm} by requiring that the $i$th components of the elements in the sequence satisfy $Q_M^{(i)}$.  The sets $X_M$ are defined as in~\refeqn{Xm}.  We additionally define
\[
X_{M}^{(i)} = \{x\in \Z_p^n \mid x_{A_{M}^{(i)}}\in Q_{M}^{(i)} \},
\]
for $i\le\sz(M)$, and $X_{M}^{(i)} = \Z_p^n$ otherwise.  Note that $X_M = \prod_{i=1}^\sz X_{M}^{(i)}$ in the sense that, for each sequence $x^{(i)}\in X_{M}^{(i)}$ with $i=1,\dots,\sz$, there is a corresponding element $x\in X_M$ with $x_j = (x_j^{(1)}, \dots, x_j^{(\sz)})$.

Now we make our choice for $Q_M^{(i)}$.  Let $U\subseteq \Z_p$ be a set with small Fourier bias and some $\delta = |U|/p$ that exists due to \refthm{FourierBias}.  We define $Q_M^{(i)}$ as consisting of all $x\in \Z_p^{A_{M}^{(i)}}$ such that the sum of the elements of $x$ belongs to $U$.  With this definition,
\begin{equation}
\label{eqn:density}
|X_M^{(i)}| = \delta p^n.
\end{equation}
Hence, there are exactly $\delta q^n$ elements $x\in\group^n$ such that $x_{A_M^{(i)}}\in T_M^{(i)}$.  If we let $\delta = 1/(2\sz|\cert|)$, a calculation similar to~\refeqn{Ysize} shows that $|Y|\ge q^n/2$.  Also, by considering each $i\in[\sz]$ independently, it is easy to see that all $X_M$ satisfy the orthogonality condition.  Thus, \reflem{GammaNorm} applies, and~\refeqn{GammaNorm} holds.

Now it remains to estimate $\|\Gamma'\|$, and it is done by considering matrix $\hGamma$ as described in \refsec{outline}, and performed once in \refsec{boundedly}.  If $\tGamma'=0$, then also $\Gamma'=0$, and we are done.  Thus, we further assume $\tGamma'\ne 0$.
\mycommand{tB}{\widetilde{B}}
\mycommand{hB}{\widehat{B}}
Recall that $(\fb_a)_{a\in\Z_p}$ denotes the Fourier basis of $\Z_p$.  The basis $e$ is defined as the Fourier basis of $\C^\group$.  It consists of the elements of the form $e_a = \bigotimes_{i=1}^\sz \fb_{a^{(i)}}$ where $a=(a^{(i)})\in\group$.  Note that $e_0$ has the required value, where $0$ is interpreted as the neutral element of $\group$.  

If $v = (v_j) = (v_j^{(i)})\in \group^n$, we define $e_v = \bigotimes_{j=1}^n e_{v_j}$, and 
$v^{(i)}\in \Z_p^n$ as $(v_1^{(i)},\dots,v_n^{(i)})$.  Also, for $w = (w_j)\in \Z_p^n$, we define
$\fb_w = \bigotimes_{j=1}^n \fb_{w_j}$.

Fix an arbitrary $M\in\cert$.  Let $\tB_M = (\tG'_M)^*\tG'_M$ and $\hB_M = (\hG_M)^*\hG_M$.  We aim to show that
\begin{equation}
\label{eqn:difference}
\|\tB_M - \hB_M\| \to 0\quad \mbox{as}\quad p\to\infty,
\end{equation}
because this implies
\[
 \|(\tGamma')^*\tGamma' - (\hGamma)^*\hGamma \| = \normB|\sum_{M\in\cert} (\tB_M - \hB_M)  | 
\le \sum_{M\in\cert} \| \tB_M - \hB_M \| \to 0
\]
as $p\to\infty$.  As $\|\tGamma'\|>0$, this implies that $\|\Gamma'\| \le 2 \|\tGamma'\|$ for $p$ large enough, and together with~\refeqn{GammaNorm} and \refthm{adv}, this implies \refthm{certificates}.

From~\refeqn{gmPrim}, we conclude that the eigenbasis of $\tB_M$ consists of the vectors $e_v$, with $v\in\group^n$, defined above.  In order to understand $\hB_M$ better, we have to understand how $e_v\elem[X_M]$ behave.  We have
\begin{equation}
\label{eqn:evInner}
(e_v\elem[X_M])^*(e_{v'}\elem[X_M]) = \prod_{i=1}^\sz (\fb_{v^{(i)}}\elem[X_{M}^{(i)}])^*(\fb_{v'^{(i)}}\elem[X_{M}^{(i)}]).
\end{equation}
Hence, it suffices to understand the behaviour of $\fb_w\elem[X_M^{(i)}]$.  For $w\in \Z_p^n$, $A\subseteq [n]$ and $c\in\Z_p$, we write $w+cA$ for the sequence $w'\in \Z_p^n$ defined by
\[
w'_j = \begin{cases}
w_j + c,& j\in A;\\
w_j,&\mbox{otherwise.}
\end{cases}
\]
In this case, we say that $w$ and $w'$ are obtained from each other by a {\em shift on $A$}.

\begin{clm}
\label{clm:zeta}
Assume $w,w'\in\Z_p^n$, and let $\xi = (\fb_w\elem[X_M^{(i)}])^*(\fb_{w'}\elem[X_M^{(i)}])$.  If $w=w'$, then $\xi = \delta$.  If $w\ne w'$, but $w$ can be obtained from $w'$ by a shift on $A_M^{(i)}$, then $|\xi|\le \fnorm|U|$.  Finally, if $w$ cannot be obtained from $w'$ by a shift on $A_M^{(i)}$, then $\xi = 0$.
\end{clm}

\pfstart
Arbitrary enumerate the elements of $U = \{u_1,\dots,u_m\}$ where $m=\delta p$.  Denote, for the sake of brevity, $A = A_M^{(i)}$.  Consider the decomposition $X_{M}^{(i)} = \bigsqcup_{k=1}^m X_k$, where
\[
X_k = \sfig{w\in \Z_p^n \mid \sum\nolimits_{j\in A} w_j = u_k} .
\]
\mycommand{ww}{\bar{w}}
Fix an arbitrary element $a\in A$ and denote $\ww = w-w_aA$ and $\ww' = w'-w'_aA$.  In both of them, $\ww_a = \ww'_a=0$, and by an argument similar to \refclm{LMi}, we get that
\begin{equation}
\label{eqn:chivInner}
(\fb_{\ww}\elem[X_k])^* (\fb_{\ww'}\elem[X_k])=
\begin{cases}
1/p,& \ww = \ww';\\
0,& \text{otherwise.}
\end{cases}
\end{equation}

If $x\in X_k$, then
\begin{multline*}
\fb_w\elem[x] = \prod_{j=1}^n \fb_{w_j}\elem[x_j] = 
\frac1{\sqrt{p^n}}\exp\skC[\frac{2\pi\ii}{p} \sum_{j=1}^n w_jx_j]
\\ = \frac1{\sqrt{p^n}}\exp\skC[\frac{2\pi\ii}{p}{\sB[ \sum_{j=1}^n \ww_j x_j + w_a\sum_{j\in A} x_j]}] = 
\exp\sB[\frac{2\pi\ii\; w_au_k}{p}]
\fb_{\ww}\elem[x] .
\end{multline*}
Hence,
\begin{equation}
\label{eqn:chivInner2}
(\fb_{w}\elem[X_M^{(i)}])^* (\fb_{w'}\elem[X_M^{(i)}])
= \sum_{k=1}^m (\fb_{w}\elem[X_k])^* (\fb_{w'}\elem[X_k])
= \sum_{k=1}^m 
\ee^{2\pi\ii (w'_a-w_a)u_k/p} 
%\exp\sB[\frac{2\pi\ii (w'_a-w_a)u_k}p ]
(\fb_{\ww}\elem[X_k])^* (\fb_{\ww'}\elem[X_k]).
\end{equation}
If $w'$ cannot be obtained from $w$ by a shift on $A$, then $\ww\ne\ww'$ and~\refeqn{chivInner2} equals zero by~\refeqn{chivInner}.  If $w=w'$, then~\refeqn{chivInner2} equals $m/p = \delta$.  Finally, if $w'$ can be obtained from $w$ by a shift on $A$ but $w\ne w'$, then $\ww=\ww'$ and $w_a\ne w'_a$.  By~\refeqn{chivInner} and~\refeqn{fnorm}, we get that~\refeqn{chivInner2} does not exceed $\fnorm|U|$ in absolute value.
\pfend

Let $v\in\group^n$, and $S=\{j\in[n]\mid v_j\ne 0\}$.  Let $v'\in\group^n$, and define $S'$ similarly.  By~\refeqn{hGM}, \refeqn{gmPrim}, \refeqn{density} and~\refeqn{evInner}, we have
\begin{equation}
\label{eqn:hBmEntry}
e_v^* \hB_M e_{v'} = \frac{q^n\beta_S(M)\beta_{S'}(M)}{|X_M|}(e_v\elem[X_M])^*(e_{v'}\elem[X_M]) = \frac{\beta_S(M)\beta_{S'}(M)}{\delta^{\sz}} \prod_{i=1}^\sz (\fb_{v^{(i)}}\elem[X_{M}^{(i)}])^*(\fb_{v'^{(i)}}\elem[X_{M}^{(i)}]).
\end{equation}
By this and \refclm{zeta}, we have that
\begin{equation}
\label{eqn:hBmDiagonal}
e_v^* \hB_M e_{v} = \beta_S(M)^2 = e_v^* \tB_M e_{v}.
\end{equation}
Call $v$ and $v'$ {\em equivalent}, if $\beta_S(M)$ and $\beta_{S'}(M)$ are both non-zero and, for each $i\in[\sz]$, $v^{(i)}$ can be obtained from $v'^{(i)}$ by a shift on $A_M^{(i)}$.  By~\refeqn{hBmEntry} and \refclm{zeta}, we have that $e_v^* \hB_M e_{v'}$ is non-zero only if $v$ and $v'$ are equivalent.

For each $i\in[\sz]$, there are at most $|A_M^{(i)}|\le n$ shifts of $v^{(i)}$ on $A_M^{(i)}$ that have an element with an index in $A_M^{(i)}$ equal to 0.  By~\refeqn{alphaZero}, the latter is a necessary condition for $\beta_S(M)$ being non-zero.  Hence, for each $v\in \group^n$, there are at most $n^\sz$ elements of $\group^n$ equivalent to it.

Thus, in the basis of $e_v$s, the matrix $\hB_M$ has the following properties.  By~\refeqn{hBmDiagonal}, its diagonal entries equal the diagonal entries of $\tB_M$, and the latter matrix is diagonal.  Next, $\hB_M$ is block-diagonal with the blocks of size at most $n^\sz$.  By~\refeqn{hBmEntry} and \refclm{zeta}, the off-diagonal elements satisfy 
\[
|e_v^*\hB_M e_{v'}| \le \frac{\fnorm|U|}{\delta} \beta_S(M)\beta_{S'}(M),
\]
because $\fnorm|U|\le\delta$. 
Since the values of $\beta_S(M)$ do not depend on $p$, and by \refthm{FourierBias}, the off-diagonal elements of $\hB_M$ tend to zero as $p$ tends to infinity.  Since the sizes of the blocks also do not depend on $p$, the norm of $\tB_M-\hB_M$ also tends to 0, as required in~\refeqn{difference}.  This finishes the proof of \refthm{certificates}.

%\section{Summary and Future Research}
%In this paper, we proved that the quantum query complexity and the learning graph complexity of a boundedly generated certificate structure are almost equal.  It is interesting to understand to what extent this result is generalizable to other certificate structures.  The difference between the learning graph complexity of the hidden shift certificate structure (\refprp{collision}) and the quantum query complexity of the hidden shift problem (as described in \refsec{examples}) may be an indication that this statement does not hold for an arbitrary certificate structure.  
%
%Also, it is possible to consider models that take the values of the variables into consideration, but not to the full extent.  For example, one may study the complexity of counting certificates.  This problem is not directly expressible in the certificate structure model.

\subsection*{Acknowledgments}
A.B. would like to thank Troy Lee, Robin Kothari and  Rajat Mittal for sharing their ideas on the limitations of learning graphs.  In particular, the notion of the learning graph complexity of a certificate structure and the proof of~\refprp{ksum} stem from these ideas.

A.B. has been supported by the European Social Fund within the project ``Support for Doctoral Studies at University of Latvia'' and by FET-Open project QCS.  A.R. acknowledges the support of Mike and Ophelia Lazaridis Fellowship and David R. Cheriton Graduate Scholarship.

%\release{
%\bibliographystyle{../../habbrvE}
%}
%\draft{
%\bibliographystyle{../../habbrv}
%}

\bibliography{bib}

\end{document}